\documentclass[12pt]{iopart}

\usepackage{natbib}
\usepackage{epsf,epsfig}
\usepackage{aasdefs,aas_macros}

\bibpunct[; ]{[}{]}{,}{n}{,}{,}

\def\beq{\begin{equation}}
\def\enq{\end{equation}}
\def\ba{\begin{eqnarray}}
\def\ea{\end{eqnarray}}
\def\bec{\begin{center}}
\def\enc{\end{center}}

\def\eps{\epsilon}

\def\siml{\lower4pt \hbox{$\buildrel < \over \sim$}}
\def\simg{\lower4pt \hbox{$\buildrel > \over \sim$}}

\def\s{{\rm s}}

\newcommand{\newblock}{}

\newcommand{\chandra}{\textit{Chandra}}

\newcommand{\bepposax}{\textit{Beppo-SAX}}

\newcommand{\swift}{\textit{Swift}}

\newcommand{\hete}{\textit{HETE-2}}
\newcommand{\integral}{\textit{INTEGRAL}}

\newcommand{\xray}{\mbox{X-ray}}

\newcommand{\Eiso}{\mbox{$E_{\gamma,{\rm iso}}$}}

\begin{document}

\title[GRB Fireball Physics]{GRB Fireball Physics: Prompt and Early Emission}

\author{D.~B. Fox \& P.~ M\'esz\'aros }
\address{Dept.\ of Astronomy \& Astrophysics and Dept.\ of Physics,
         Pennsylvania State University,
         University Park, PA 16802, USA}

\begin{abstract}
We review the fireball shock model of gamma-ray burst prompt and early
afterglow emission in the light of rapid follow-up measurements made
and enabled by the multi-wavelength \swift\ satellite. These
observations are leading to a reappraisal and expansion of the
previous standard view of the GRB and its fireball. New information on
the behavior of the burst and afterglow on minutes to hour timescales
has led, among other results, to the discovery and follow-up of short
GRB afterglows, the opening up of the $z\simg 6$ redshift range, and
the first prompt multi-wavelength observations of a long
GRB-supernova.  We discuss the salient observational results and some
associated theoretical issues.
\end{abstract}

\maketitle


\section{Observational Advances in the Swift Era}
\label{sec:obs}

NASA's \swift\ mission \cite{gcg+04} has enabled fundamental insights
into the physics of gamma-ray bursts thanks to two new capabilities:
First, the greater sensitivity of its Burst Alert Detector
\cite{bbc+05} (BAT; energy range 20--150 keV) in comparison to the
preceding \bepposax\ and \hete\ missions \cite{band06}; and second,
its ability to slew in less than 100 seconds to the burst direction
determined by the BAT, which allows it to position its much
higher-angular resolution \xray\ (XRT, few-arcsec) and UV-Optical
(UVOT, sub-arcsec) detectors \cite{bhn+05,rhm+04} for observations of
the prompt and early afterglow emission.


As of July~2006, over 150 bursts had been detected by BAT, at an
average rate of 2 bursts detected per week. Of these, roughly 90\%
were followed promptly with the XRT within $350$ s from the trigger,
and about half within 100~s \cite{brg+05}, while $\sim 30\%$ were
detected with the UVOT \cite{rsf+05}.  This resulted in over 30 new
redshift determinations.  Eleven definitive short GRBs were detected,
of which seven had detected \xray\ afterglows (including three with
optical afterglows, and two with radio), and five had proposed
redshifts.


These \swift\ observations, complemented by the continuing operations
of the \hete\ and \integral\ missions, have brought the total number
of redshift determinations to over 60 since \bepposax\ enabled the
first one in 1997. The redshifts based on \swift\ have a median $z\sim
2.8$ \cite{jlf+06,bmb+06}, which is a factor $\simg 2$ higher than the
median of those previously culled via \bepposax\ and \hete\
\cite{bkf+05}. This can be credited largely to the prompt $\sim$arcsec
positions from XRT and UVOT, which make rapid ground-based
observations possible at a stage when the afterglow is still bright.
The highest \swift-enabled redshift so far was that of GRB\,050904,
from 8-meter class Subaru spectroscopy, $z=6.29$ \cite{kka+05} (the
pre-\swift\ record was $z=4.5$ for the IPN-localized GRB\,000131
\cite{ahp+00}). The relative paucity of UVOT detections versus XRT
detections may be ascribed in part to this higher median redshift (and
correspondingly reduced median flux), and in part to the (expected)
increased dust extinction at the shorter rest-frame wavelengths
implied for any given observing band \cite{rsf+05}, although the issue
is still the subject of debate.


{\it BAT light curves:} 
The BAT is provided with an array of triggering algorithms which
provide greater sensitivity than those of previous spacecraft,
including ``imaging'' triggers which, through analysis of the counts
integrated over 128-s to 512-s timescales, enable the detection of
faint, slow-rising events.  In a peculiar irony of GRB studies, these
image triggers have made possible the discovery of both the farthest,
highly time-dilated burst GRB\,050904, and the nearest, faint and
slow-rising burst GRB\,060218 (see below for further discussion).  For
some of the bursts which fall in the ``long'' category ($\gamma$-ray
duration $t_\gamma \simg 2$ s), the BAT has detected faint soft
gamma-ray tails which extend the duration by a factor up to two beyond
what the previous Burst and Transient Source Experiment (BATSE) could
have detected \cite{gehrels06}. Such extended soft tails have been
found also in some ``short'' bursts, normally defined as having a hard
spectrum and duration $t_\gamma \siml 2$ s (see below).

\begin{figure}[tbh]
\centerline{\epsfxsize=5.in \epsfbox{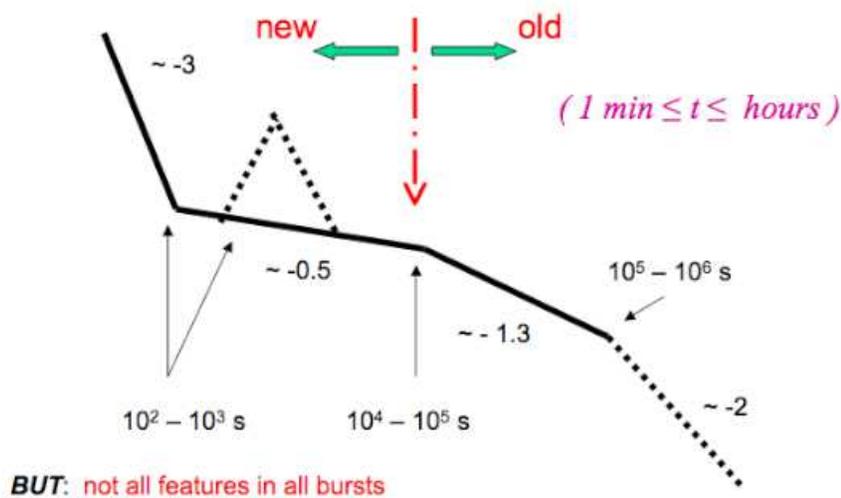}}
\caption{Schematic features seen in early \xray\ afterglows detected with the
\swift\ XRT instrument (e.g.\ \cite{zfd+06,nkg+06}); see text.
\label{fig:xrt-lc}}
\end{figure}

{\it XRT light curves:}
Striking new insights into burst and afterglow physics have come from
the detailed \xray\ light curves, starting on average 100 seconds
after the $\gamma$-ray trigger, that result from the prompt XRT
observations of BAT-detected bursts.  These observations suggest a
canonical \xray\ afterglow (Fig.~\ref{fig:xrt-lc};
\cite{nkg+06,zfd+06}) with one or more of the following stages (note
that the numerical subscript enumerates each stage):
(1) An initial steep decay $F_X \propto t^{-\alpha_1}$ with temporal
index $3 \siml \alpha_1 \siml 5$, and an energy spectrum $F_\nu
\propto \nu^{-\beta_1}$ with energy spectral index $1 \siml \beta_1
\siml 2$ (photon index $0\siml \Gamma\siml 1$), extending up to a time
$300\,\s \siml t_1 \siml 500\,\s$;
(2) A subsequent flatter decay $F_X \propto t^{-\alpha_2}$ with $0.2
\siml \alpha_2 \siml 0.8$ and energy index $0.7 \siml \beta_2 \siml
1.2$, at times $10^3\,\s \siml t_2 \siml 10^4\,\s$ (in some
cases interspersed with flares, see \cite{zfd+06};
(3) A ``normal'' decay $F_X \propto t^{-\alpha_3}$ with $1.1 \siml
\alpha_3 \siml 1.7$ and $0.7 \siml \beta_3 \siml 1.2$ (generally
unchanged from the previous stage, i.e.\ $\beta_3\approx \beta_2$), up
to a time $t_3 \sim 10^5\,\s$, or in some cases longer;
(4) In some cases, a steeper \xray\ decay at the end, $F_X \propto
t^{-\alpha_4}$, with $2\siml \alpha_4 \siml 3$, after $t_4\sim 10^5
\,\s$, resembling what is expected from jet breaks;
(5) In about half the afterglows, one or more \xray\ flares or bumps
are observed in the light curve, sometimes starting as early as 100 s
after trigger, and sometimes as late as $10^5\,\s$.  The energy in
these flares ranges from a percent up to a value comparable to the
prompt emission (in GRB\,050502b).  The rise and decay times of these
flares is unusually steep, depending on the reference time $t_0$,
behaving as $(t-t_0)^{\pm \alpha_{\rm fl}}$ with $3 \siml \alpha_{\rm
  fl} \siml 6$, and energy indices which can be also steeper than
during the smooth decay portions. The flux level after the flare
usually decays to the value extrapolated from the value before the
flare rise.

\begin{figure}[tbh]
\centerline{\epsfxsize=5.in \epsfbox{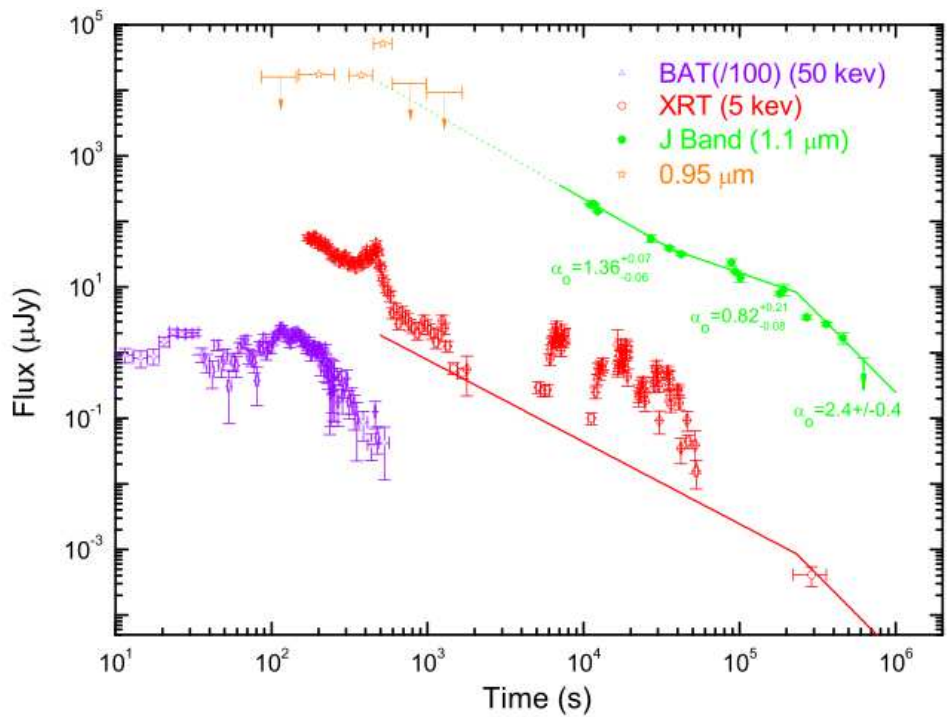}}
\caption{Prompt and afterglow lightcurves for the redshift $z=6.29$
  GRB\,050904 at hard \xray\ (BAT, 50~keV), \xray\ (XRT, 5~keV), and
  near-infrared ($J$-band, 1.1~\micron; and $I$-band, 9500\,\AA)
  frequencies.  A bright early flare is seen in the IR and \xray; in
  the \xray\ this is followed by extensive flaring activity for more
  than a day after the burst.  Power-law decay indices measured in the
  IR ($\alpha_{\rm o}$) are indicated, where available; a likely jet
  break is seen at $t\approx 2.3$\,days \cite{tac+05}.  Data from
  \cite{hnr+06,bad+06,cmc+06,tac+05}.}
\label{fig:lc050904}
\end{figure}

{\it Very high-redshift bursts:}
Another major advance achieved by \swift\ was the detection of the
long burst GRB\,050904, which broke the $z>6$ redshift barrier. This
burst was very bright, both in its prompt $\gamma$-ray emission
($\Eiso\sim 10^{54}$ erg) and in its \xray\ afterglow. Prompt
ground-based optical/IR upper limits and a $J$-band detection
suggested a photometric redshift $z>6$ \cite{hnr+06}. Spectroscopic
confirmation with the 8.2 m Subaru telescope gave $z=6.29$
\cite{kka+05}.  There are several striking features to this burst.
One is its enormous \xray\ brightness, exceeding for a full day the
\xray\ brightness of the most distant quasar known to-date,
SDSS~J0130+0524 -- and exceeding it by a factor of $10^5$ in the first
minutes \cite{wrh+06}. The implications for the use of GRBs as a tool
for probing the IGM are thought-provoking. Another notable feature was
its extremely variable \xray\ light curve, showing many large
amplitude flares throughout the first day (Fig.~\ref{fig:lc050904}). A
third exciting feature is its brief, very bright IR flash
\cite{bad+06}, comparable in brightness to the famous $m_V\sim 9$
optical flash in GRB\,990123.

%
%
%
%

{\it Short bursts:}
The third major advance from \swift\ was the discovery and
localization of short GRB afterglows. As of July~2006, eleven
definitive short bursts have been localized by \swift, and \hete\ has
discovered two short bursts with subsequent afterglow emission; in
addition, the short burst GRB\,051103 was localized by the IPN to a
region intersecting the nearby galaxies M81 and M82.  In seven of the
\swift\ short bursts and two \hete\ bursts an \xray\ afterglow was
detected and observed, with GRBs 050709, 050724, 051221A, 060313, and
060121 (as well as the candidate short burst GRB\,051227) also showing
optical afterglows, and GRBs 050724 and 051221A being detected in the
radio \cite{berger06}.  These are the first afterglows detected for
short bursts, and yielded the first host galaxy identifications.  The
hosts are of early type (ellipticals) in roughly half the cases, and
otherwise appear to be dwarf irregular galaxies -- albeit with
evidence for old as well as young stellar populations
\cite{sbk+06}. The redshifts of five of them are in the range 0.15 to
0.5, while GRB\,050813 was initially thought to have $z=0.7$
\cite{pbc+06} but may be associated with a galaxy cluster at $z\approx
1.8$ \cite{berger06}, and the redshift of the \hete\ GRB\,060121 is
estimated to be $z\approx 4.6$ \cite{dcg+06}. The median $z$ (as yet
unaffected by these two relatively high redshifts) is $z_{\rm
med}=0.26$, which gives a $(1+z)$ that is 1/3 that of the long bursts.
While there is evidence for star formation in roughly half the host
galaxies, overall the host properties correspond to expectations for a
progenitor population of neutron star or neutron star-black hole
binary systems, the most often-discussed progenitor candidates.  Such
hosts would also be compatible with other progenitors involving old
compact stars.

The first short burst afterglow seen by \swift\ was that of
GRB\,050509B, a low luminosity ($\Eiso\sim 2\times 10^{48}$ erg) burst
with a simple power-law \xray\ afterglow which could only be followed
for $\sim 10^4$ s \cite{gso+05}.  The subsequent afterglows of
GRB\,050724 ($\Eiso\sim 3\times 10^{50}$ erg; \cite{bcb+05}) and
GRB\,051221A ($\Eiso\sim 1.5\times 10^{51}$ erg; \cite{sbk+06}) were
brighter, and could be followed in \mbox{X-rays} for at least $10^5$ s
\cite{gbp+06,bgc+06}.  In particular, the \xray\ afterglow of
GRB\,050724 is remarkable in that it exhibits a typical \xray\ light
curve as observed from long GRBs by \swift\ (apart from the absence of
a slow-decay phase).  It shows a fast early decay, with significant
\xray\ flares, one at 100~s and another at $3\times 10^4$~s. The first
flare has the same fluence as the prompt emission, while the late
flare has roughly 10\% of that. The interpretation of this activity
poses interesting challenges, as discussed below.


{\it The GRB-SN connection:}
The fourth major advance from \swift\ was its observation, with BAT,
XRT and UVOT, of an unusually long ($\sim 2000$ s), soft burst,
GRB\,060218 \cite{cmb+06}, soon found to be associated with SN2006aj,
a nearby ($z=0.033$) type Ic supernova
\cite{msg+06,mha+06,pmm+06,skn+06}.  The optical light curve of this
supernova peaked earlier than for most other known supernovae, and the
time of core collapse is constrained to within a day of the GRB
trigger.  This was the first time that a connected GRB and supernova
event was observed starting in the first $\sim$100~s in \xray\ and
UV/optical light, and the results are of great interest. The early
\xray\ light curve shows a slow rise and plateau followed by a drop
after $\sim 10^3$ s; the spectrum is initially dominated by a
power-law component, with an increasing thermal component that
dominates after $\approx$3000~s.  Perhaps the most interesting
interpretation involves shock break-out of a semi-relativistic
component in a WR progenitor wind \cite{cmb+06}, although see also
\cite{fp06}.  After this a more conventional \xray\ power-law decay
follows, and a UV component peaking at a later time can be interpreted
as due to the slower supernova envelope shock.  Finally, a
radio-bright component is seen that presents an interesting comparison
to other local and GRB-related type Ic supernovae \cite{skn+06}.  A
distinct GRB/SN detection based on \swift\ afterglow observations is
that associated with GRB\,050525A \cite{dmb+06}.

%
%
%

{\it Prompt optical observations:}
A final major advance enabled by \swift\ has been a wealth of
information about the prompt and early-time optical emission of GRBs.
Thanks to \swift\ alerts and an array of ground-based robotic
telescopes (as well as the UVOT), the detection of optical
counterparts during and shortly after ($<$100~s) the gamma-ray trigger
is becoming almost routine.  The \swift\ era began with the detection
in optical \cite{vww+05} and near-infrared \cite{bbs+05} wavelengths
of a bright flash associated with GRB\,041219A (localized in real time
by \integral).  Apart from exhibiting the first optical flash with
brightness (after correcting for Galactic extinction) similar to
GRB\,990123, this burst showed the first evidence for correlated low-
and high-energy emission, intepreted as optical emission from the same
internal shocks that produced the GRB \cite{vww+05,bbs+05}.  A similar
level of correlated optical emission was later observed from
GRB\,050820A \cite{vww+06}; however in this case it was accompanied by
an uncorrelated component that connected smoothly to the later
afterglow.  Prompt optical emission that extrapolates smoothly to the
later afterglow seems to be the more usual case; however, bright
flares consistent with a reverse-shock interpretation are seen on
occasion \cite{jpk+06,kgs+06}.  The most exciting single observation
is probably the $I=14.1$\,mag optical flash from GRB\,050904 at
$z=6.3$ observed with TAROT \cite{bad+06}, which rises to its peak
after the end of gamma-ray emission and is brighter, and decays more
quickly than, the later afterglow -- all properties consistent with an
origin in this burst's reverse shock.


\section{Prompt Emission Models}
\label{sec:prompt}

The prompt gamma-ray emission of a classical GRB has most of its
energy in the energy range 0.1 to 2.0~MeV. The generic photon spectrum
is a broken power law \cite{bmf+93} with a break energy in the above
range, and, typically, power law extensions down into the X-ray, and
up into the 100 MeV to GeV ranges (although a number of bursts show a
thermal-like spectrum, which sometimes can be fitted with a blackbody
\cite{ryde04}).


\subsection{Gamma-Ray Emission}
\label{sub:prompt:gamma}

The simplest model for the gamma-ray emission assumes
that a proton crossing a strong shock front with a relative bulk
Lorentz factor $\Gamma_{21}$ acquires (in the comoving frame) an
internal energy characterized by a random (comoving) Lorentz factor
$\gamma_{p,m} \sim \Gamma$ \cite{mr93a}. The comoving magnetic field
behind the shock can build up due to turbulent dynamo effects behind
the shocks \cite{mr93a,mr93b} (as also inferred in supernova remnant
shocks). More recently, the Weibel instability has been studied in
this context \cite{ml99,msf+04,spitkovsky05}.  The efficiency of this
process remains under debate, but one can parametrize the magnetic
field as having a post-shock energy density which is a fraction
$\eps_B$ of the equipartition value relative to the proton random
energy, $B'\sim [32 \pi \eps_B n_{ex} (\gamma'_p -1)m_p c^2]^{1/2}
\Gamma$, where the post-shock proton comoving internal energy is
$(\gamma'_p-1) m_p c^2 \sim ~1$ (or $\sim \Gamma $) for internal
(external) shocks \cite{mr93a,rm94}.  Scattering of electrons (and
protons) by magnetic irregularities upstream and downstream can lead
to a Fermi acceleration process resulting in a relativistic power law
distribution of energies $N(\gamma)\propto \gamma^{-p}$ with $p\geq
2$.  The starting minimum (comoving) Lorentz factor of the thermal
electrons injected into the acceleration process, $\gamma_{e,m}$ would
in principle be the same as for the protons, $\Gamma$ (they
experience the same velocity difference), hence both before and after
acceleration they would have $\sim (m_e/m_p)$ less energy than the
protons. However, the shocks are collisionless, i.e.\ mediated by
chaotic electric and magnetic fields, and can redistribute the proton
energy between the electrons and protons, up to some fraction $\eps_e$
of the thermal energy equipartition value with the protons, so
$\gamma_{e,m} \sim \eps_e (m_p/m_e) \Gamma$ \cite{mr93b,mrp94}. The
synchrotron spectrum, in its simplest form, peaks at $\nu_m \sim
\Gamma (3/8\pi)(eB'/m_e c)\gamma_m^2\sim 2\times 10^6 B'\gamma_m^2
\Gamma ~{\rm Hz}$, and has a shape $F_\nu \propto [ \nu^{1/3};~
\nu^{-(p-1)/2}]$ for [$\nu< \nu_m$;~ $\nu>\nu_m$], in the adiabatic
limit where the cooling time is longer than the dynamic time. In the
opposite (radiative) limit, the spectrum peaks at the cooling
frequency $\nu_c$ (at which cooling time for electrons emitting at
$\nu_c$ equals dynamic time, and the spectrum above is $F_\nu\propto
\nu^{-p/2}$. Depending on the parameters, more complicated spectra are
possible, discussed e.g.\ in \cite{spn98}. The synchrotron spectrum is
modified at low energies by self-absorption \cite{mr93b,gps99b},
making the spectrum steeper; and it is also modified at high energies,
due to inverse Compton effects \cite{mr93b,mrp94,dcm00}, extending
into the GeV range.

However, a number of effects can modify the simple synchrotron
spectrum.  For instance, the distribution of observed low energy
spectral indices $\beta_1$ (where $F_\nu\propto \nu^{\beta_1}$ below
the spectral peak) has a mean value $\beta_1\sim 0$, but for a
fraction of bursts this slope reaches positive values $\beta_1>1/3$
which are incompatible with a simple synchrotron interpretation
\cite{pbm+00}. Possible explanations include synchrotron
self-absorption in the \xray\ \cite{gps00} or in the optical range
up-scattered to X-rays \cite{pm00}, low-pitch angle scattering or
jitter radiation \cite{medvedev00,medvedev06}, observational selection
biases \cite{lp00} and/or time-dependent acceleration and radiation
\cite{lp02}, where low-pitch angle diffusion can also explain high
energy indices steeper than predicted by isotropic scattering. Other
models invoke a photospheric component and pair formation
\cite{mr00b}.  Pair formation can become important in internal shocks
or dissipation regions occurring at small radii, since a high comoving
luminosity implies a large comoving compactness parameter. A moderate
to high scattering depth can lead to a Compton equilibrium which gives
spectral peaks in the right energy range \cite{pw04,pw05}. An
important aspect is that Compton equilibrium of internal shock
electrons or pairs with photospheric photons leads to a high radiative
efficiency, as well as to spectra with a break at the right preferred
energy and steep low energy slopes \cite{rm05,pmr05,pmr06a}. It also
leads to possible physical explanations for the Amati \cite{aft+02} or
Ghirlanda \cite{ggl04} relations between spectral peak energy and
burst fluence \cite{rm05,thompson06}.


\subsection{Optical Emission}
\label{sub:prompt:optical}

Prompt optical flashes, defined as optical emission detected while
gamma-ray emission is still in progress and exemplified by the
original detection from GRB\,990123 \cite{abb+99}, have been generally
interpreted \cite{sp99a,mr99,np05} as radiation from the reverse
component of the external shock. However, a prompt optical flash can
be produced from either an internal shock or the reverse external
shock, or both \cite{mr97a,mr99}.  The decay rate of the optical flux
from reverse shocks is very fast (and that of internal shocks, faster
yet) compared to the decay of forward shock emission, so that the
forward shock component typically dominates within minutes to tens of
minutes.

Observations prior to the \swift\ era demonstrated already that bright
flashes such as that seen from GRB\,990123 were relatively rare. At
the same time, the slow-fading ($\alpha\sim 0.4$) optical emission
from GRB\,021004 and the relatively faint optical flash from
GRB\,021211 were both argued to have properties consistent with a
reverse shock origin \cite{kz03a,zkm03,fyk+03,fps+03,lfc+03,wei03}.

Optical/UV afterglows are now detected at early times, $<$100~s, in
prompt observations with the \swift\ UVOT telescope in roughly half
the bursts for which an \xray\ afterglow was seen; for a detailed
discussion see \cite{rsf+05}.  Of particular interest is the ongoing
discussion of whether the ``dark GRBs'' which remain undetected by the
UVOT are really optically deficient, or remain unobserved due to
observational biases (e.g., the relatively blue bandpass of UVOT and
the high redshifts of the \swift\ sample; \cite{bkf+05}).

The fastest routine responses to \swift\ alerts are being realized by
robotic ground-based telescopes, a large number of which have been
brought on-line in recent years.  The first substantial discovery
yielded by these observatories has been of the gamma-ray correlated
component of the prompt optical emission \cite{vww+05,bbs+05,vww+06}.
This component is not observed in every burst, but the mere fact of
this correlation is sufficient to establish its likely origin in the
burst's internal shocks \cite{mr97a,mr99}.  When observed, the ratio
of the correlated gamma-ray to optical flux densities has been found
to be roughly $10^5$ to one.

In contrast to bursts with reverse-shock flare or gamma-ray correlated
emission, the typical burst is now revealed to either exhibit a single
power-law decay from early times \cite{ryk+05,qry+06,yar+06}, or to
exhibit a flat or rising light curve \cite{rsp+04,rmy+06} before it
enters the standard power-law afterglow decay.  The initial brightness
of the typical counterpart is $V\sim 14$ to 17~mag, which has made
observations challenging for the usual $<$1~m telescopes employed.

There are a number of possible reasons for the observed faintness of
early optical emission from typical GRBs. Suppression of internal
shock emission can be provided by self-absorption in the optical,
coupled to the lower flux implied by the $\nu^{1/3}$ low-energy
asymptote of the synchrotron spectrum peaking at $\sim$ MeV
\cite{mr97a}. Suppression of reverse shock emission, on the other
hand, may indicate the absence or weakness of the reverse shock, e.g.\
if the ejecta are highly magnetized \cite{mr97a}. Alternatively, the
deceleration might occur in the thick-shell regime ($T \gg t_{dec}$),
resulting in the reverse shock being relativistic and boosting the
optical spectrum into the UV \cite{kobayashi00} (but a detection by
UVOT might then be expected, unless the decay is faster than the
typical 100 to 200~s UVOT response time). Another possibility, for a
high comoving luminosity, is copious pair formation in the ejecta,
causing the reverse shock spectrum to peak in the IR
\cite{mrr+02}. Both GRBs 990123 and 050904 (see below) have a
relatively large $\Eiso\sim 10^{54}$\,erg, so the latter seems a
promising option.  Even without pairs, more accurate calculations of
the reverse shock \cite{np04,mkp06} find the emission to be
significantly weaker than what was estimated earlier. Yet another
possibility is that the cooling frequency in reverse shock is not much
larger than the optical band frequency.  In this case the optical
emission from the reverse shock drops to zero very rapidly soon after
the reverse shock has crossed the ejecta and the cooling frequency
drops below the optical and there are no electrons left to radiate in
the optical band \cite{mkp06}.

That said, a few persuasive observations of reverse-shock optical
emission have been made in the \swift\ era, so it can finally be said
that GRB\,990123 does not stand alone.  The early optical/IR emission
from GRB\,041219 would have rivalled that seen from GRB\,990123 if not
for the large Galactic extinction along the line of sight
\cite{vww+05}, and of the three distinct peaks observed by PAIRITEL,
the second may represent a reverse shock contribution \cite{bbs+05} --
if so, a relatively small Lorentz factor, $\Gamma\siml 70$, is
implied.  Observations of GRB\,050525A with UVOT \cite{bbb+06} and
GRB\,060111B with TAROT (in a unique time-resolved tracking mode;
\cite{kgs+06}) show the ``flattening'' light-curve familiar from
GRB\,021211 (as well as, probably, GRB\,990123) that is termed the
``type II'' light curve by \cite{zkm03}.  Evolution of the optical
flux in this manner is supposed to indicate the presence of magnetized
ejecta or a Poynting-flux dominated outflow.

An alternate case may be provided by GRB\,060117, observed by the FRAM
sky monitor telescope of the Pierre Auger Observatory \cite{jpk+06}.
This burst is the largest-fluence GRB observed by \swift\ to-date, and
its optical flash was also very bright, peaking at $R\approx 10$\,mag
and thus rivalling GRB\,990123.  Although their quality is not high,
the data suggest that the forward shock peak is distinguished above
the decay of the reverse shock flux, providing a potential first
example of the ``type I'' two-peaked lightcurve that is predicted for
a hydrodynamic (nonmagnetized) outflow \cite{zkm03}.  

Finally, the most exciting prompt robotic IR detection (and optical
non-detection) is that of GRB\,050904 \cite{bad+06,hnr+06}, which is
also though to be due to a reverse shock \cite{fzw05,wyf06} (see,
however, \cite{wei06}).  This object, at the unprecedented ``very
high'' redshift of $z=6.29$ \cite{kka+05}, had an \xray\ brightness
exceeding for a day that of the brightest \xray\ quasars
\cite{wrh+06}, and its optical/IR brightness in the first 500~s
(observer time) was comparable to that of GRB\,990123, with a
similarly steep time-decay slope $\alpha\sim 3$.


\section{Early Afterglow Models}
\label{sec:afterglow}

The afterglow generally becomes important after a time
\beq
t_{ag}= {\rm Max}[  (r_{dec}/2c \Gamma^2)(1+z)~,~T] =
{\rm Max}[ 10^2 (E_{52}/n_0)^{1/3} \Gamma_2^{-8/3}(1+z)|{\rm s}~,~T]~,
\label{eq:tag}
\enq
where the deceleration time is $t_{dec}\sim (r_{dec}/2c \Gamma^2)$ and
$T$ is the duration of the prompt outflow; $t_{ag}$ then marks the
beginning of the self-similar blast-wave regime where $\Gamma\propto
r^{-3/2}\propto t^{-3/8}$ (in the adiabatic regime; $\Gamma\propto
r^{-3}\propto t^{-3/7}$ in the radiative regime).

Denoting the frequency and time dependence of the afterglow spectral
energy flux as $F_\nu(t)\propto \nu^{-\beta}t^{-\alpha}$, the late
\xray\ afterglow phases (3) and (4) described above are similar to
those known previously from \bepposax\ (for a review of this earlier
behavior and its modeling see e.g.\ \cite{zm04}). The ``normal'' decay
phase (3), with temporal decay indices $\alpha \sim 1.1-1.5$ and
spectral energy indices $\beta\sim 0.7-1.0$, is what is expected from
the evolution of the forward shock in the Blandford-McKee self-similar
late time regime, under the assumption of synchrotron emission.

The late steep decay phase (4) of \S \ref{sec:obs}, occasionally seen
in \swift\ bursts, is naturally explained as a jet break, when the
decrease of the ejecta Lorentz factor leads to the light-cone angle
becoming larger than the jet angular extent, $\Gamma_j(t) \simg
1/\theta_j$ (e.g.\ \cite{zm04}). It is noteworthy, however, that this
final steepening has been seen in less than $\sim 10\%$ of the \swift\
afterglows, and even then for the most part only in X-rays. The
corresponding optical light curve breaks have been few, and not
well-constrained.  This is unlike the case with the $\sim 20$
\bepposax\ bursts, for which many achromatic breaks were reported in
the optical \cite{fks+01}, while in some of the rare cases where an
\xray\ or radio break was reported it occurred at a different time
\cite{bkf03}. The relative paucity of optical breaks in \swift\
afterglows may be an observational selection effect due to the larger
median redshift, and hence fainter and redder optical afterglows at
the same observer epoch, as well as perhaps reluctance to commit large
telescope time on the more frequently-reported \swift\ bursts (an
average, roughly, of two per month with \bepposax\ versus two per week
with \swift).


\subsection{Steep Decay}
\label{sub:steep}

Among the new early afterglow features detected by \swift, the steep
initial decay phase $F_\nu \propto t^{-3}- t^{-5}$ in X-rays of the
long GRB afterglows is one of the most puzzling. There could be
several possible reasons for this. The most immediate of these would
be the cooling following cessation of the prompt emission (internal
shocks or dissipation). If the comoving magnetic field in the emission
region is random [or transverse], the flux per unit frequency along
the line of sight in a given energy band, as a function of the
electron energy index $p$, decays as $F_\nu \propto t^{-\alpha}$ with
$\alpha={-2p}~[(1-3p)/2]$ in the slow cooling regime, where
$\beta=(p-1)/2$, and it decays as $\alpha=-2(1+p),~[-(2-3p)/2]$ in the
fast cooling regime where $\beta=p/2$, i.e.\ for the standard $p=2.5$
this would be $\alpha=-5,~[-3.25]$ in the slow cooling or
$\alpha=-7,~[-2.75]$ in the fast cooling regime, for random
[transverse] fields \cite{mr99}. In some bursts this may be the
explanation, but in others the time and spectral indices do not
correspond well.

Currently the most widely considered explanation for the fast decay,
either in the initial phase (1) or in the steep flares, attributes it
to off-axis emission from regions at $\theta >\Gamma^{-1}$ (also termed
curvature effects or high-latitude emission \cite{kp00}). In this
case, after line-of-sight gamma-ray emission has ceased, the off-axis
emission observed from $\theta>\Gamma^{-1}$ is $(\Gamma\theta)^{-6}$
smaller than that from the line of sight. Integrating over the equal
arrival time region, this flux ratio becomes $\propto
(\Gamma\theta)^{-4}$. Since the emission from $\theta$ arrives
$(\Gamma\theta)^2$ later than from $\theta=0$, the observer sees the
flux falling as $F_\nu\propto t^{-2}$, if the flux were frequency
independent. For a source-frame flux $\propto \nu'^{-\beta}$, the
observed flux per unit frequency varies then as
\beq
    F_\nu\propto (t-t_0)^{-2-\beta}
\label{eq:hilat}
\enq
i.e.\ $\alpha=2+\beta$. This high-latitude radiation, which for
observers outside the line cone at $\theta > \Gamma^{-1}$ would appear
as prompt $\gamma$-ray emission from dissipation at radius $r$,
appears to observers along the line of sight (inside the light cone)
to arrive delayed by $t\sim r\theta^2/2c)$ relative to the trigger
time, and its spectrum is softened by a Doppler factor $D\propto
t^{-1}$ into the \xray\ observer band.  For the initial prompt decay,
the onset of the afterglow (e.g.\ phases 2 or 3), which also comes
from the line of sight, may overlap in time with the delayed
high-latitude emission.  In equation (\ref{eq:hilat}) $t_0$ can be
taken as the trigger time, or some value comparable or less than by
equation (\ref{eq:tag}). This can be used to constrain the prompt
emission radius \cite{lb06}. When $t_{dec}<T$, the emission can have
an admixture of high-latitude and afterglow contributions, and since
the afterglow has a steeper spectrum than the high-latitude emission
(which has a prompt spectrum), one can have steeper decays
\cite{owo+06}. Values of $t_0$ closer to the onset of the decay also
lead to steeper slopes. Structured jets, when viewed on-beam, produce
essentially the same slopes as homogeneous jets, while off-beam
observing can lead to shallower slopes \cite{dzf05}.  For the flares,
if their origin is assumed to be internal (e.g.\ some form of late
internal shock or dissipation) the value of $t_0$ is just before the
flare, near the observed time of flare onset \cite{zhang06}.  This
interpretation appears, so far, compatible with most of the \swift\
afterglows \cite{zfd+06,nkg+06,pmg+06}.

Alternatively, the initial fast decay could be due to the emission of
a cocoon of exhaust gas \cite{pmr06b}, where the temporal and spectral
index are explained through an approximately power-law behavior of
escape times and spectral modification of multiply-scattered photons.
The fast decay may also be due to the reverse shock emission, if
inverse Compton interactions up-scatter the (primarily synchrotron)
optical photons into the \xray\ range. The decay starts after the
reverse shock has crossed the ejecta and electrons are no longer
accelerated, and may have both line-of-sight and off-axis components
\cite{kzm+06}.  This poses strong constraints on the Compton $y$
parameter, and cannot explain steeper decays with $\alpha>2$, or
$\alpha>2+\beta$ if the off-axis contribution dominates. Models
involving bullets -- whose origin, acceleration and survivability is
unexplained -- could give a prompt decay index $\alpha=3$ to 5
\cite{ddd06}, but imply a bremsstrahlung energy index $\beta \sim 0$
which is not observed in the fast decay, and require
fine-tuning. Finally, a ``patchy shell'' model, where the Lorentz
factor is highly variable in angle, would produce emission with
$\alpha\sim 2.5$. Thus, such mechanisms may explain the more gradual
decays, but not the more extreme $\alpha=5$ to 7 values encountered in
some cases.


\subsection{Shallow Decay}
\label{sub:shallow}

The slow decay portion of the \xray\ light curves ($\alpha\sim 0.3$ to
0.7), quite ubiquitously detected by \swift, is not entirely new,
having been detected in a few cases by \bepposax. This, as well as the
appearance of wiggles and flares in the \xray\ light curves several
hours after the burst, were the motivation for the ``refreshed
shocks'' scenario \cite{rm98,sm00}. Refreshed shocks can flatten the
afterglow light curve for hours or days, even if the ejecta is all
emitted promptly at $t=T \siml t_\gamma$, but with a range of Lorentz
factors, say $M(\Gamma) \propto \Gamma^{-s}$, where the lower $\Gamma$
shells arrive much later to the foremost fast shells which have
already been decelerated.  Thus, for an external medium of density
$\rho\propto r^{-g}$ and a prompt injection where the Lorentz factor
spread relative to ejecta mass and energy is $M(\Gamma)\propto
\Gamma^{-s}$, $E(\Gamma)\propto \Gamma^{-s+1}$, the forward shock flux
temporal decay is given by \cite{sm00}
\beq 
    \alpha=[(g-4)(1+s)+\beta(24 -7g+sg)]/[2(7+s-2g)]~.
\label{eq:shallow}
\enq
It needs to be emphasized that in this model all the ejection can be
prompt (e.g.\ over the duration $\sim T$ of the gamma ray emission)
but the low $\Gamma$ portions arrive at (and refresh) the forward
shock at late times, which can range from hours to days. That is, it
is not the central engine which is active late; rather, its effects
are seen late.  Fitting such refreshed-shock models to the shallow
decay phases in \swift\ bursts \cite{gk06} leads to a $\Gamma$
distribution which is a broken power law, extending above and below a
peak around $\sim 45$.

An alternate version of refreshed shocks does envisage central engine
activity extending for long periods of time, e.g.\ $\siml$ day (in
contrast to the $\siml$ minutes engine activity in the model
above). Such long-lived activity may be due to continued fall-back
into the central black hole \cite{wh06} or to a magnetar wind
\cite{zm01}.  One characteristic of both types of refreshed models is
that after the refreshed shocks stop and the usual decay resumes, the
flux level shows a step-up relative to the previous level, since new
energy has been injected.

From current analyses, the refreshed shock model is generally able to
explain the flatter temporal \xray\ slopes seen by \swift, both when
it is seen to join smoothly on the prompt emission (i.e.\ without an
initial steep decay phase) or when seen after an initial steep
decay. Questions remain concerning the interpretation of the fluence
ratio in the shallow \xray\ afterglow and the prompt gamma-ray
emission, which can reach $E_{\rm X}/E_\gamma\siml 1$
\cite{owo+06}. This requires a higher radiative efficiency in the
prompt gamma-ray emission than in the \xray\ afterglow. One could
speculate that this might be achieved if the prompt outflow were
Poynting-dominated.  Alternatively, a highly-efficient afterglow might
emit a large fraction of its energy in other (unobserved) bands, e.g.\
in the GeV or IR. Or \cite{ity+05} a previous mass ejection might have
emptied a cavity into which the ejecta moves, leading to greater
efficiency at later times, or otherwise causing the energy fraction
going into electrons to increase $\propto t^{1/2}$.


\subsection{X-ray Flares}
\label{sub:flares}

Refreshed shocks can also explain some of the \xray\ flares whose rise
and decay slopes are not too steep. However, this model encounters
difficulties with the very steep flares with rise or decay indices
$\alpha\sim \pm(5--7)$, such as inferred from the giant flare of
GRB\,050502b \cite{brf+05} around 300\,s after the trigger. Also, the
flux level increase in this flare is a factor $\sim 500$ above the
smooth afterglow before and after it, implying a comparable energy
excess in the low- versus high-$\Gamma$ material. An explanation based
on inverse Compton scattering in the reverse shock \cite{kzm+06} can
explain a single flare at the beginning of the afterglow, as long as
the subsequent flux decay is not too steep.  For multiple flares,
models invoking blastwave interaction with a lumpy external medium
have generic difficulties explaining steep rises and decays
\cite{zfd+06}, although extremely dense, sharp-edged lumps, if they
exist, might satisfy the steepness criterion \cite{dermer06}.

Currently the more widely-considered model for the flares ascribes
them to late central engine activity \cite{zfd+06,nkg+06,pmg+06}.  The
strongest arguments in favor of this are that the energy budget is
more easily satisfied, and the fast rises and decays more
straightforward to explain.  In such a model the flare energy can be
comparable to the prompt emission, and the fast rise comes naturally
from the short time variability leading to internal shocks (or to
rapid reconnection), while the rapid decay may be explained as the
high-latitude emission following the flare, with $t_0$ reset to the
beginning of each flare (see discussion in \cite{zhang06}).
Considering the full population of \xray\ flares, some are
well-modeled by refreshed forward shocks, while in others this is
clearly ruled out and a central engine origin is better suited
\cite{wdw+06}. Aside from the phenomenological desirability based on
energetics and timescales, a central engine origin is conceivable,
within certain time ranges, based on numerical models of the core
collapse scenario for long bursts. These invoke the core collapse of a
massive stellar progenitor, where continued infall into the
fast-rotating core can continue for a long time \cite{wh06}. However,
large flares with a fluence which is a sizable fraction of the prompt
emission occurring hours later remain difficult to understand. It has
been argued that gravitational instabilities in the infalling debris
torus can lead to lumpy accretion \cite{paz06}.  Alternatively, if the
accreting debris torus is dominated by magneto-hydrodynamic (MHD)
effects, magnetic instabilities can lead to extended, highly
time-variable accretion \cite{pb03}.


\subsection{Short Burst Afterglows}
\label{sub:short}

\swift\ and to a lesser extent \hete\ have provided the first bonafide
short burst afterglows, with the first observations beginning (in some
cases) within $<$100~s of the trigger, leading in turn to arcsec and
sub-arcsec localizations, host galaxy identifications, and redshifts
(although, as yet, none via direct absorption spectroscopy).

In the first short burst afterglow, GRB\,050509b \cite{gso+05}, the
extrapolation of the prompt BAT emission into the \xray\ range, along
with the XRT light curve from 100 to 1000~s, can be fitted with a
single power law of $\alpha \sim 1.2$.  The \xray\ coverage was sparse
due to orbital constraints and the faintness of the afterglow, and the
number of detected \xray\ photons was small. No optical transient was
detected, but an elliptical host galaxy was identified at $z=0.225$
(e.g.\ \cite{berger06}). In GRB\,050709 an optical transient was
identified, as well as a host galaxy \cite{ffp+05}, an irregular
galaxy at $z=0.16$ (observations also ruled out any supernova
association).

GRB\,050724 was relatively bright, and in addition to an \xray\
afterglow observed with the XRT and \chandra\ \cite{bcb+05}, yielded
decaying optical and radio afterglows \cite{bpc+05}.  This burst, as
with roughly half the short burst afterglows seen to-date, is
associated with an elliptical host galaxy. It also had a
low-luminosity, soft gamma-ray extension of the short-hard gamma-ray
component (which would have been missed by BATSE), and it had an
interesting \xray\ afterglow extending beyond $10^5$~s, with no jet
break seen to the limits of the final \chandra\ observation
\cite{gbp+06}. The soft gamma-ray extension, lasting up to 200~s,
connects smoothly with the beginning of the XRT afterglow, which has
$\alpha\sim 2$ between 100 and 300~s, and then enters a much steeper
decay with $\alpha\sim 5$ to 7 out to $\sim 600 s$, followed by a more
moderate decay $\alpha \sim 1$. An unexpected feature is a strong
flare peaking at $5\times 10^4$~s, with an energy 10\% that of the
prompt emission, and an amplitude which represents a 10-fold increase
over the preceding slow decay.

GRB\,051221A, the next panchromatic short burst afterglow, has an
extended light curve in the \xray\ \cite{bgc+06} and optical
\cite{sbk+06}.  This was the brightest short burst observed by \swift\
to date, and occurred in a star-forming galaxy at redshift $z=0.5464$.
The \xray\ light curve also provides the clearest evidence of
``standard'' energy injection, with a flattening from 1.4 to 3.4 hours
after the burst which increases the afterglow energy by a factor of
$\sim$3, and is moreover reflected in contemporaneous radio
detections from the VLA.  

With seven afterglows and five relatively secure redshifts in hand via
host galaxy identifications, the distribution of short bursts in
redshift and among host galaxy types -- including an equal number of
spiral/irregular and elliptical hosts -- is typical of an old
($\simg$Gyr) population of progenitors, such as neutron star (NS)
binaries or black hole-neutron star binaries \cite{ngf05}.

The main challenges posed by the short burst afterglows are the
relatively long, soft tail of the prompt emission, and the strength
and late occurrence of the flares. A possible explanation for the
extended long soft tails ($\sim$100~s) may be that the compact binary
progenitor is a black hole - neutron star system \cite{bcb+05}, since
analytical and numerical arguments (\cite{dlk05}, and references
therein) suggest that disruption of the NS and its disappearance into
the black hole may lead to a complex and extended accretion phase
significantly longer than for double neutron stars. The flares, for
which the simplest interpretation might be as refreshed shocks
(compatible with a short engine duration $T\siml t_\gamma \sim 2$ s,
for ejecta with an extended Lorentz factor distribution), require the
energy in the slow material to be at least ten times more energetic
than the fast material responsible for the prompt emission in the
specific case of the GRB\,050724 flare at $10^4$ s. The rise and decay
times are moderate enough for this interpretation. Another
interpretation invokes the accretion-induced collapse of a neutron star
in a binary, leading to a flare when the fireball created by the
collapse hits the companion \cite{mrz06}, which might explain moderate
energy one-time flares.  However, for repeated, energetic flares, as
also seen from long bursts, the total energetics are easier to satisfy
if one postulates late central engine activity (lasting at least half
a day), containing $\sim 10\%$ of the prompt fluence \cite{bcb+05}. A
possible way to produce this might be via temporary ``choking'' of an
MHD outflow \cite{pb03} (c.f.\ \cite{vo01}), which might imply linear
polarization of the \xray\ flare \cite{fzp05}. Such MHD effects could
plausibly also explain the initial $\sim$100~s soft tail. However, a
justification for substantial $\simg$10$^5$~s features remains so far
on tentative grounds.

The similarity of the \xray\ afterglow light curves to those of long
bursts is, in itself, an argument in favor of the prevalent view that
the afterglows of both long and short bursts can be described by the
same paradigm, independently of any differences in the
progenitors. This impression is reinforced by the fact that the \xray\
light curve temporal slope is, on average, that expected from the
usual forward shock afterglow model, and that in two short bursts (so
far) there is evidence for what appears to be a jet break
\cite{ffp+05,bgc+06,sbk+06}. However, while very similar, the
first-order differences are revealing: the average isotropic energy of
the short bursts is a factor of $\sim$100 smaller, while the average
jet opening angle (based on two breaks and one lower limit) is a
factor of $\sim 2$ larger \cite{ffp+05,gbp+06,bgc+06,sbk+06}.  Using the
standard afterglow theory, the bulk Lorentz factor decay can be
expressed through $\Gamma(t_d) =6.5(n_o/E_{50})^{1/8} t_d^{-3/8}$,
where $t_d=(t/{\rm day})$, $n_o$ is the external density in units of
cm$^{-3}$, and $E_{50}$ is the isotropic equivalent energy in units of
$10^{50}$\,erg.  If the jet break occurs at
$\Gamma(t_{br})=\theta_j^{-1}$ the jet opening angle and the total jet
energy $E_j$ are
\beq 
    \theta_j= 9^o (n_o/E_{50})^{1/8} t_{d,br}^{3/8}~~, 
    E_j = \pi \theta_j^2 E \sim 10^{49} n_o^{1/4} (E_{50}
          t_{d,br})^{3/4}~{\rm erg}~. 
\label{eq:short}
\enq
For the first three well-studied afterglows, GRBs 050709, 050724, and
051221A, these equations, together with the standard afterglow
expressions for the flux level as a function of time before and after
the break, lead to fits \cite{panaitescu06,bgc+06,sbk+06} which are
not completely determined, allowing for GRBs 050709 and 051221A either
a very low or a moderately low external density, and for GRB\,050724 a
moderately low to large external density. The main uncertainties are
in the jet break times, which are poorly sampled, and in the absence
of high-quality radio data for any burst (GRB\,050724 had a reasonable
radio flux, but too much variability in all bands).  More bright short
bursts like GRB 051221A will be needed to improve the jet break
statistics substantially.


\subsection{The Supernova Component}
\label{sub:supernova}

GRB-associated supernovae have been regularly observed as ``red
bumps'' in the light curves of low-redshift GRBs, and more rarely, by
direct spectroscopic observation at 7 to 30 days after the GRB.

The observation of the GRB-supernova GRB\,060218/SN\,2006aj with
\swift, however, has raised the prospect of an entirely new component
of emission at early times, namely, the shock breakout of the
simultaneous supernova explosion.  The thermal component of the prompt
XRT emission, which evolves into the UVOT range over the course of the
first few hours after the burst, has been interpreted as the shock
break-out of the GRB/SN in an optically-thick wind of a Wolf-Rayet
progenitor star \cite{cmb+06} (but see also \cite{fp06}).  It is
important that the wind be sufficiently dense; given the known
redshift of this burst, the blackbody radius of the early XRT/UVOT
thermal component is roughly 100 times the size of a Wolf-Rayet star.
At the same time, however, the absence of hydrogen features (e.g.\
H$\alpha$) in the SN spectrum requires such a progenitor, and
disallows a red giant progenitor, which would produce a type II SN.

The observation of a single such event in one year of \swift\
observations offers the tantalizing prospect of gathering more
examples with each additional year of \swift\ operations.  In
particular, intensive ground-based observations tracking the evolution
of this thermal component to near-infrared, millimeter, and radio
wavelengths will be useful in confirming or refuting the
shock-breakout interpretation.


\bigskip
\noindent
{\it Acknowledgements:} 
We are grateful to the \swift\ team for collaborations and to NSF AST
0307376 and NASA NAG5 13286 for support.  We thank Lijun Gou for
preparation of Figure~\ref{fig:lc050904}.

\bigskip





\end{document}